# Achromatic super-oscillatory lenses with sub-wavelength focusing

Guang Hui Yuan[1,#], Edward T. F. Rogers[2,3], and Nikolay I. Zheludev[1,2*]

[1]Centre for Disruptive Photonic Technologies, The Photonic Institute, SPMS, Nanyang Technological University, Singapore 637371, Singapore

[2]Optoelectronics Research Centre and Centre for Photonic Metamaterials, University of Southampton, Highfield, Southampton, SO17 1BJ, UK

[3]Instiute for Life Sciences, University of Southampton, Highfield, Southampton, SO17 1BJ, UK

Corresponding authors: #ghyuan@ntu.edu.sg and *nzheludev@ntu.edu.sg

**Abstract:** Lenses are crucial to light-enabled technologies. Conventional lenses have been perfected to achieve near-diffraction-limited resolution and minimal chromatic aberrations. However, such lenses are bulky and cannot focus light into a hotspot smaller than half wavelength of light. Pupil filters, initially suggested by Toraldo di Francia, can overcome the resolution constraints of conventional lenses, but are not intrinsically chromatically corrected. Here we report single-element planar lenses that not only deliver sub-wavelength focusing – beating the diffraction limit of conventional refractive lenses – but also focus light of different colors into the same hotspot. Using the principle of super-oscillations we designed and fabricated a range of binary dielectric and metallic lenses for visible and infrared parts of the spectrum that are manufactured on silicon wafers, silica substrates and optical fiber tips. Such low cost, compact lenses could be useful in mobile devices, data storage, surveillance, robotics, space applications, imaging, manufacturing with light, and spatially resolved nonlinear microscopies.

## INTRODUCTION

Chromatic aberration and the resolution limit are two major challenges for high-performance optical imaging. Chromatic aberration (chromatism) is a failure of the lens to focus all colours to the same point. In refractive focusing devices, it results from dispersion of the material from which the lens is made from. In diffractive focusing devices, it results from the accumulated wavelength-dependent phase delay of electromagnetic waves forming the focus. To reduce chromatic aberrations, complex optical elements such as achromatic doublet, triplet and diffractive-refractive hybrid lenses have been built[1,2], made with components of opposite dispersion properties[3]. However, such lenses are inevitably bulky which complicates integration. Several approaches have been proposed to miniaturize beam-shaping and focusing devices including planar Fresnel zone plates and, more recently, metasurface-based plasmonic and dielectric lenses and axicons[4-12]. However, chromatism remains a key challenge, which has been tackled by the use of dielectric metasurfaces[13,14], wavelength-independent geometric phase[15] and by exploiting the inherent dispersion in diffractive optics[16].

The other key parameter of an imaging system is its spatial resolution: the ability to resolve details of the object that is being imaged. It is commonly believed that resolution of optical system that images object from the far-field to the far-field of the lens is limited to half the optical wavelength (the Abbe-Rayleigh diffraction limit), due to the loss of fine details of the electromagnetic field distribution near the object. Indeed, the Abbe-Rayleigh diffraction limit was seen as the main obstacle for development of sub-wavelength-resolution label-free microscopy. Powerful alternative approaches for super-resolution fluorescent bio-imaging have been developed. These include stimulated emission depletion microscopy (STED) which uses two beams, one to excite and the other to deplete luminescence[17], photo-activated localization microscopy (PALM) and stochastic optical reconstruction microscopy (STORM) based on localising single luminescent molecules in the imaged object[18,19]. A common feature of these techniques is the use of fluorescent labels embedded in the object that make them suitable only for a narrow group of applications, predominantly in biology.

Many important technological challenges, from next generation lithography to data storage and fabrication with light, could be solved if a focusing system existed that could beat the Abbe-Rayleigh diffraction limit and create hotspots much smaller than the wavelength of light. The existence of the resolution limit has been challenged by the promising idea of the far-field to the far-field superlens fabricated from a material with negative index of refraction[20]. Such a superlens could "translate" the both propagating waves and evanescent waves in the immediate proximity of the object into the remote image. Although more simple devices imaging from the near-field to the near-field[21] and from the near-field to the far-field[22,23] have been successfully demonstrated, the negative index superlens that images from the far-field to the far-field has not yet been developed for the optical part of the spectrum and would likely inherit chromatic aberration from the dispersion of the negative index medium.

As originally proposed by Toraldo di Francia, focusing beyond the Abbe-Rayleigh diffraction limit can be achieved by pupil filtering technique[24-27]. From the prospective of modern wave theory this approach exploits the phenomenon of optical super-oscillations[28-38] that implies the non-trivial fact that complex band-limited signal can locally oscillate much faster than its highest Fourier components components[39-41] and thus accurately tailored interference of waves can form foci smaller than size allowed by the Abbe-Rayleigh diffraction limit. Mathematically, according to M.V. Berry this is explainable by the fact that "In the Wigner representations of the local Fourier transform in the 'phase space' Wigner function can have both positive and negative values, which causes subtle cancellations in the Fourier integration over all of the function"[42]. In super-oscillatory focusing and imaging devices, fine details of the electromagnetic field near the object are conveyed to the image by the propagating waves themselves, the effect which the conventional Abbe-Rayleigh theory deemed impossible since spatial spectrum of free-space waves is limited by the free-space wavevector.

This work is devoted to the development of achromatic super-oscillatory sub-wavelength focusing devices. We explore planar masks that convert light waves into super-oscillatory foci by the virtue of tailored interference of waves diffracted from different areas of the mask (termed as 'super-oscillatory lens', SOL). To create super-oscillatory foci, both amplitude and phase masks can be used. Although the outcome of interference is wavelength-dependent, an SOL can be designed to focus different wavelengths into the same spot (see Fig. 1). This is possible because SOL can create foci of extremely long depth, extending tens of wavelengths away from the mask[33,43,44], so foci of different wavelengths can partially overlap creating a zone of distances from the SOL where a range of colors can be focused simultaneously. An alternative approach is to use super-oscillatory mask generating a number of discrete foci at different distances from it. For different wavelengths some of these foci can overlap and thus the SOL will focus two or more wavelengths simultaneously in one spot.

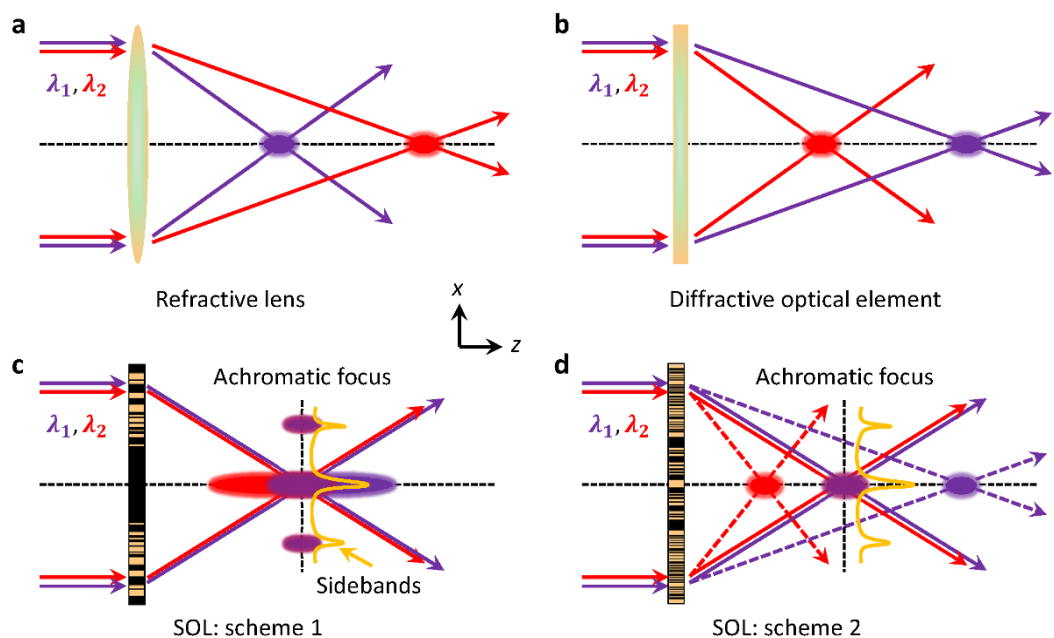

**Figure 1** Achromatic super-oscillatory lens (SOL) schemes and comparison with conventional refractive lens and diffractive optical element. (**a**) Dispersion of refractive lens (larger focal length for longer wavelength). (**b**) Opposite dispersion of diffractive optical element (larger focal length for shorter wavelength). (**c**) Optical SOL for achromatic focusing: two long depth-of-focus foci overlap to form the achromatic focus. (**d**) SOL focuses two wavelengths $\lambda_1$ and $\lambda_2$ into multiple discrete hotspots along the axial direction that spatially overlap to achieve the bi-chromatic focusing.

Below we will describe a range of SOLs, demonstrating the wide variety of lenses that can be designed and the flexibility possible in designing a lens for particular applications. We have made both amplitude- and phase-modulated achromatic SOLs for visible and infrared radiation and an apochromatic red/green/blue SOL for the visible part of the spectrum. Here and below we adopt the well-established terminology[1-3] that a lens is called achromatic if it brings two wavelengths into focus in the same plane and apochromatic if it focuses three wavelengths simultaneously.

## MATERIALS AND METHODS

**Achromatic SOL design procedure.**

For design of achromatic SOLs we used the particle swarm optimization (PSO) algorithm[45], which optimizes a problem with regard to a given merit function using a population of 'particles' in the $N$-dimensional search space. For a ring mask, the radial direction is divided into $N$ equally spaced zones. Each zone has either unit ('1') or zero ('0') value which corresponds to transparent and opaque for amplitude mask, and two phase levels for phase mask. The PSO algorithm searches for the best arrangement of these binary values. The target function to describe the electric field intensity profiles near focus for individual wavelength $\lambda_p$ is defined as

$$I_p^{tar}(\lambda_p, r, z) = \left[\frac{2J_1(a_p r)}{a_p r}\right]^2 \exp\left[-\frac{(z-z_f)^2}{b_p^2}\right] \qquad (1)$$

where $p$ is the numbering of wavelengths, $J_1(x)$ is the first-order Bessel function of the first kind, $z_f$ is the desired achromatic working distance, $a_p = \frac{3.233}{\text{FWHM}_p}$, $b_p = \frac{\text{DOF}_p}{2\sqrt{ln2}}$, $\text{FWHM}_p$ is the full-width half maximum of transverse hotspot size and $\text{DOF}_p$ is the depth of focus at $\lambda_p$. The merit function to achieve the achromatic SOL design is given as

$$F(r,z) = \sum_{\lambda_p} \left|I_p^{act}(\lambda_p, r, z) - I_p^{tar}(\lambda_p, r, z)\right|^2 \qquad (2)$$

where $I_p^{act}(\lambda_p, r, z)$ is the normalized actual intensity distribution at $\lambda_p$ and is calculated by angular spectrum method for a given mask design. The optimal SOL mask is achieved after sufficient iterations that improvement in $F(r,z)$ saturates.

**SOL sample preparation and fabrication.**

All SOLs were fabricated by focused ion beam milling (FIB, FEI Helios Nanolab 650). For the binary amplitude mask, a 100 nm-thick gold film was deposited on a silica glass substrate using a thermal evaporator (Oerlikon Univex 250) with a deposition rate of 0.2 Å/s. Prior to deposition of gold, a 5 nm-thick chromium adhesion layer was deposited on the substrate. For the fiberized SOL the fabrication process remains essentially the same, while the photonic crystal fibers (LMA35, NKT photonics) was cleaved with a large-diameter fiber cleaver (VYTRAN LDC-400). A careful alignment between the centers of SOL and fiber core had to be achieved for correct performance of the SOL. For the dielectric SOL, FIB writing is conducted directly onto the silicon wafer (University Wafer Inc., DSP, 100, p-type).

## RESULTS AND DISCUSSION
**Near-IR achromatic fiberized amplitude mask SOL**

A sub-wavelength focusing, near-IR lens on a fiber tip would be valuable in many applications, most notably imaging through silicon wafers and substrates inside silicon chips, diagnostics of optoelectronic devices, high-resolution two-photon polymerization nano-fabrication, non-destructive imaging of in-vitro biomedical samples, micro-spectrometry of molecular vibrational modes, and as an interconnect element of silicon photonic devices. We choose the tip of a single-mode large-mode-area photonic crystal fiber (PCF, NKT photonics) as the platform for our achromatic near-infrared SOL. The advantage of such fibers is that they provide a stable high-quality mode, even at high output power, and single-mode operation with almost constant mode diameter for broadband wavelengths – which is useful, though not essential – in designing achromatic devices.

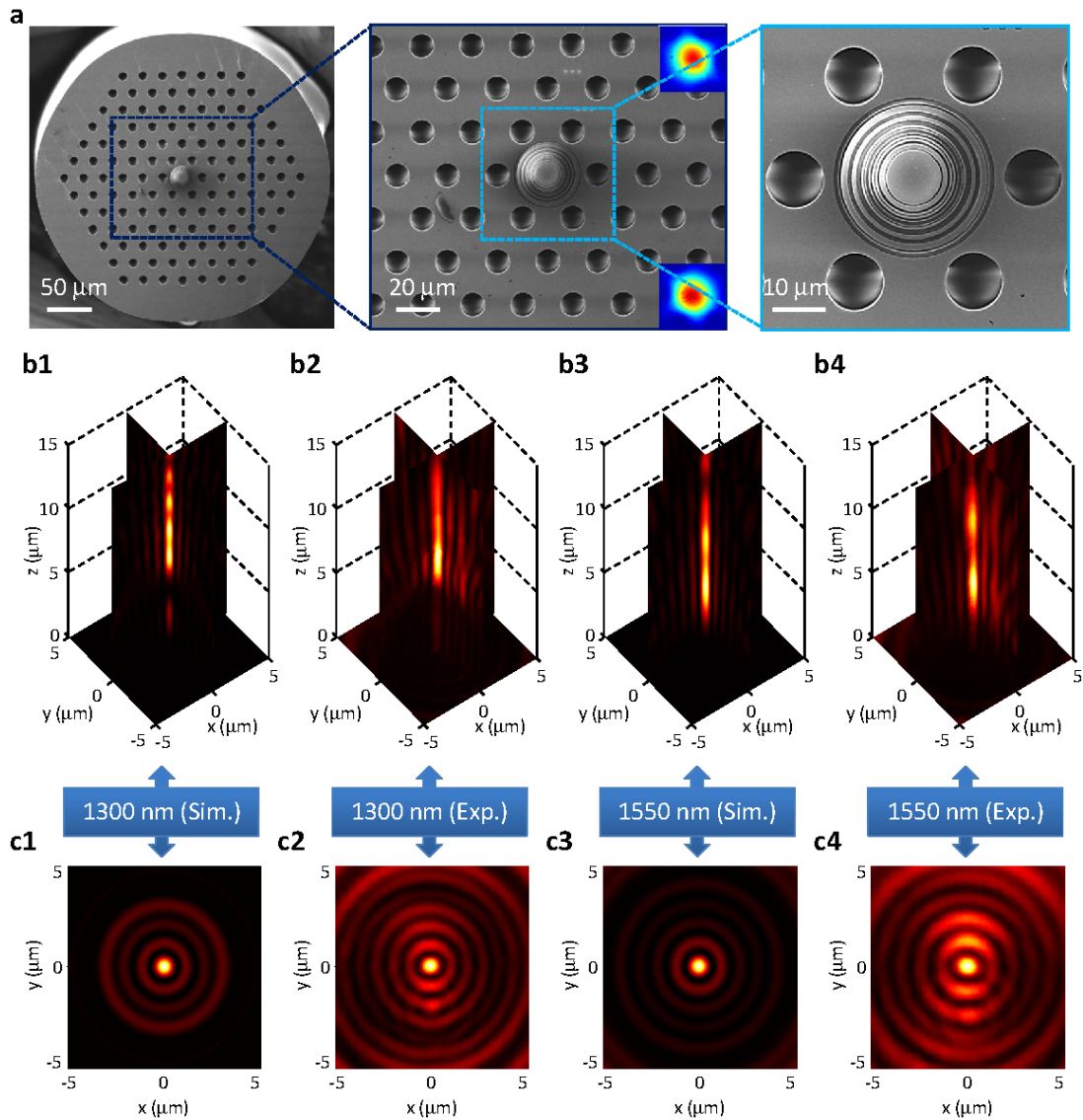

**Figure 2** Near-IR achromatic fiberized amplitude-mask SOL. (**a**) SEM image and zoomed-in views of the SOL manufactured on the core of a single-mode large-mode-area photonic crystal fiber where a 100 nm-thick gold film was deposited on the fiber end after cleaving. Insets in the middle figure show the fiber modes at the wavelengths of $\lambda_{IR1}$=1.3 µm (top) and $\lambda_{IR2}$=1.55 µm (bottom). (**b**) Experimental and simulated field patterns in two typical cross-sections, where the achromatic super-oscillatory hotspots are expected at $z_f$ =8 µm. (**c**) Corresponding intensity profiles in the achromatic transverse focal plane.

The SOL is a concentric ring nanostructure located in the middle of the PCF cross-section (See SEM micrograph in Fig. 2a). It is optimized to be achromatic at two telecommunication wavelengths $\lambda_{IR1}$=1.3 µm and $\lambda_{IR2}$=1.55 µm. The design and optimization procedures, as well as the fabrication process, are described in the Methods section below (See Supplementary Information-Section 1 for the design parameters of all the SOLs used in this paper). The theoretical focusing performance for the chosen wavelengths – calculated by angular spectrum method[30] – is shown in Figs. 2b1 and 2b3. Calculations reveal that at $\lambda_{IR1}$=1.3 µm the SOL produces an "optical needle", a long focal spot extending from axial distance of 6.1 µm to 11.9 µm, while at $\lambda_{IR2}$=1.55 µm the "optical needle" extends from 3.1 µm to 9.7 µm from the lens. Here achromatic performance can be expected from 6.1 µm to 9.7 µm centered at working distance of approximately 8 µm.

To experimentally characterize the lens, we first measured mode field diameter of the fiber and found it to be 26 µm for both wavelengths of interest, $\lambda_{IR1}$=1.3 µm and $\lambda_{IR2}$=1.55 µm, as shown in the inset of Fig. 2a. A conventional glass lens with diameter of 26 µm and focal distance of 8 µm would have a numerical aperture of NA=0.85. Since the super-oscillatory field structure of the lens is formed by free-space waves, it can be imaged with a lens of similar or higher numerical aperture. The lens was characterized with a tunable supercontinuum laser source coupled into the fiber and a high-resolution InGaAs camera equipped with an objective of NA=0.95 (See Supplementary Information-Section 2 for the experimental details). The intensity pattern after the SOL was recorded slice by slice with an axial step of 200 nm (see Figs. 2b2 and 2b4). The optical needles are located between 5.4 µm to 10.8 µm for $\lambda_{IR1}$ and between 3.4 µm and 10.4 µm for $\lambda_{IR2}$, showing reasonable agreement with simulations. The fraction of energy focused into the sub-wavelength hotspots at the target plane at 8 µm was found to be 2.9% and 2.1% for $\lambda_{IR1}$ and $\lambda_{IR2}$ respectively.

The cross-sections of the hotspots at 8 µm from the lens are presented in Fig. 2c. At $\lambda_{IR1}$=1.3 µm the full width at half maximum (FWHM) in the simulation was 0.47×$\lambda_{IR1}$ compared to (0.51±0.02)×$\lambda_{IR1}$ observed experimentally. At $\lambda_{IR2}$=1.55 µm the FWHM of the predicted hotspot was 0.44×$\lambda_{IR2}$ compared to (0.48±0.02)×$\lambda_{IR2}$ observed experimentally. A normal glass lens of the equivalent size would not be able to deliver such sharp focus: its focal spot would be at least the Abbe-Rayleigh diffraction limit of 0.59$\lambda$.

**Near-IR achromatic silicon phase-mask SOL**

Similar achromatic focusing can be achieved with dielectric phase mask. Low-loss dielectric phase masks can simultaneously improve the throughput efficiency of the lens and dramatically enhance optical breakdown thresholds of the device for both CW and pulsed illumination. We chose to work with silicon as it is compatible with the fabrication process of well-established technologies. Here we used focused ion beam milling, although high resolution optical and electron beam lithography would give similar or even better results. By milling different depths into silicon wafer we create a lossless dielectric achromatic SOL for the same telecommunication wavelengths of $\lambda_{IR1}$=1.3 µm and $\lambda_{IR2}$=1.55 µm. The refractive indices at $\lambda_{IR1}$ and $\lambda_{IR2}$ are measured to be 3.509 and 3.481 respectively (See Supplementary Information-Section 3 for the n-k curves) and material's dispersion is taken into account during the mask optimization. At a milling depth of 312 nm, a step in the silicon layer creates a phase retardation of $1.2\pi$ at $\lambda_{IR1}=1.3$ µm and $\pi$ at $\lambda_{IR2}=1.55$ µm. The SEM micrograph of an SOL is shown in Fig. 3a.

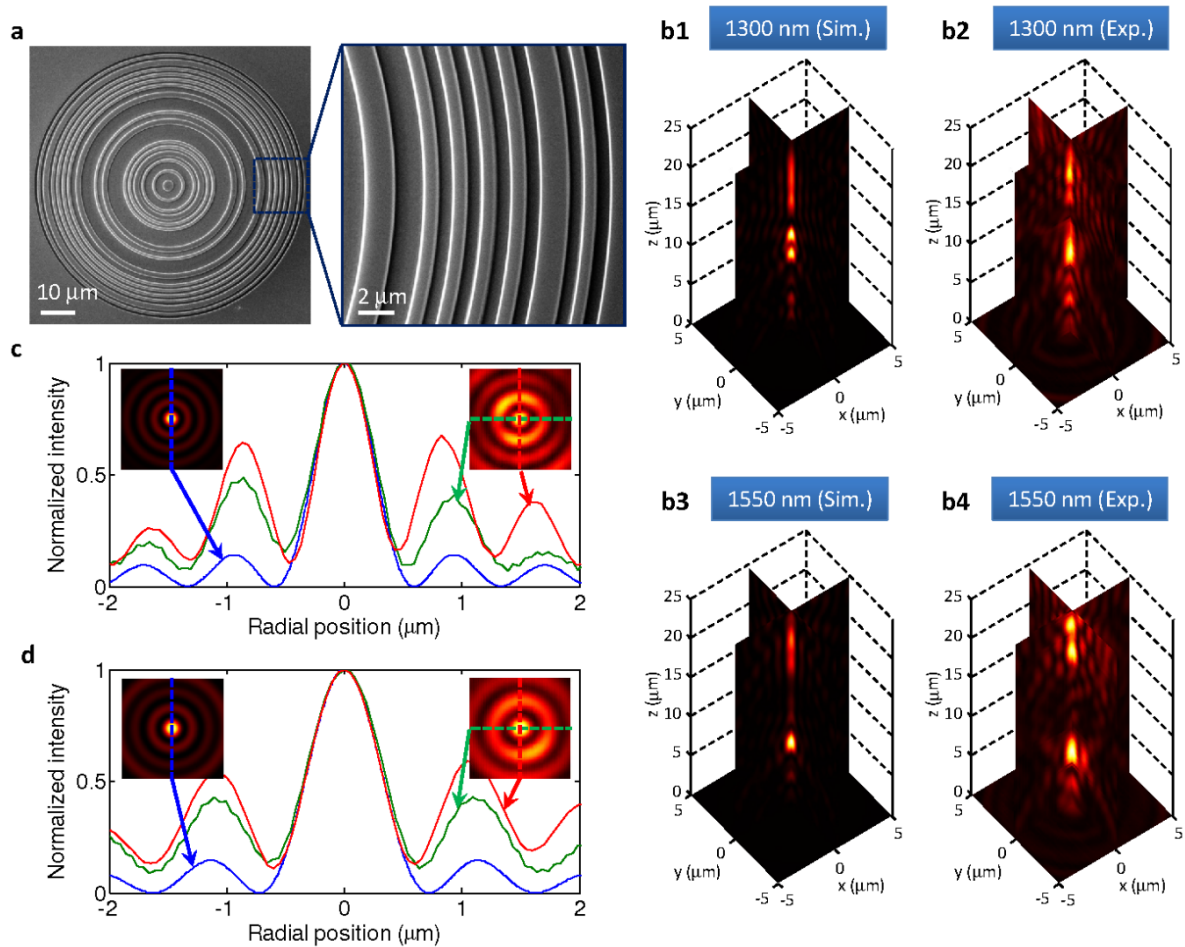

**Figure 3** Near-IR achromatic silicon phase mask SOL. (**a**) SEM micrograph of achromatic SOL milled on a silicon wafer, designed for achromatic focusing for wavelengths of $\lambda_{IR1}=1.3$ µm and $\lambda_{IR2}=1.55$ µm with working distance of $z_f=20$ µm. The right zoomed-in view illustrates the fabrication quality. (**b**) Experimental and simulated achromatic focusing behaviors in the planes parallel (*xz*) and perpendicular (*yz*) to the incident polarization. (**c**)(**d**) Comparison of the hotspot size between simulation and experiment for $\lambda_{IR1}$ (c) and for $\lambda_{IR2}$ (d). Insets are intensity profiles at $z_f=20$ µm. Blue curves are data from angular spectrum simulation. Red and green curves show the line-scan profiles perpendicular and parallel to incident polarization (*x*) respectively.

The computed and experimentally measured focusing performance of the lens is presented in Fig. 3b. The super-oscillatory hotspots can be observed near an axial distance of 20 µm, where at $\lambda_{IR1}$ the theoretical depth of focus is 6.7 µm, compared with experimental value of 5.2 µm. At $\lambda_{IR2}$ we obtained 5.4 µm and 5.8 µm for the computed and experimentally observed depths of focus. At 20 µm from the lens the hotpot size at $\lambda_{IR1}=1.3$ µm is $0.427\times\lambda_{IR1}$ (simulation, blue curve) and $(0.42\pm0.04)\times\lambda_{IR1}$ (experiment), see Fig. 3c. Similarly, for $\lambda_{IR2}$, the FWHM of the hotspot is 0.435×$\lambda_{IR2}$ (simulation) and (0.44±0.02)×$\lambda_{IR2}$ (experiment), see Fig. 3d. The fraction of energy focused into the sub-wavelength hotspots at the target plane at 20 µm was found to be 1.26% and 2.1% for $\lambda_{IR1}$ and $\lambda_{IR2}$ respectively.

**Visible achromatic amplitude mask SOL**

The interference-based super-oscillatory focusing is scalable to any wavelength. An SOL working at a shorter wavelength does put more demand of the finesse and accuracy of fabrication, but we show here that it is still possible to manufacture visible-wavelength achromatic SOLs by focused ion beam milling. As examples, we have chosen two wavelengths of $\lambda_1$=690 nm and $\lambda_2$=870 nm, one in the visible and the other in the infrared bands. These or similar wavelengths are often used in nonlinear optical imaging techniques and spectroscopic pump-probe experiment, in particular with the wide-spread Ti:Sapphire laser sources.

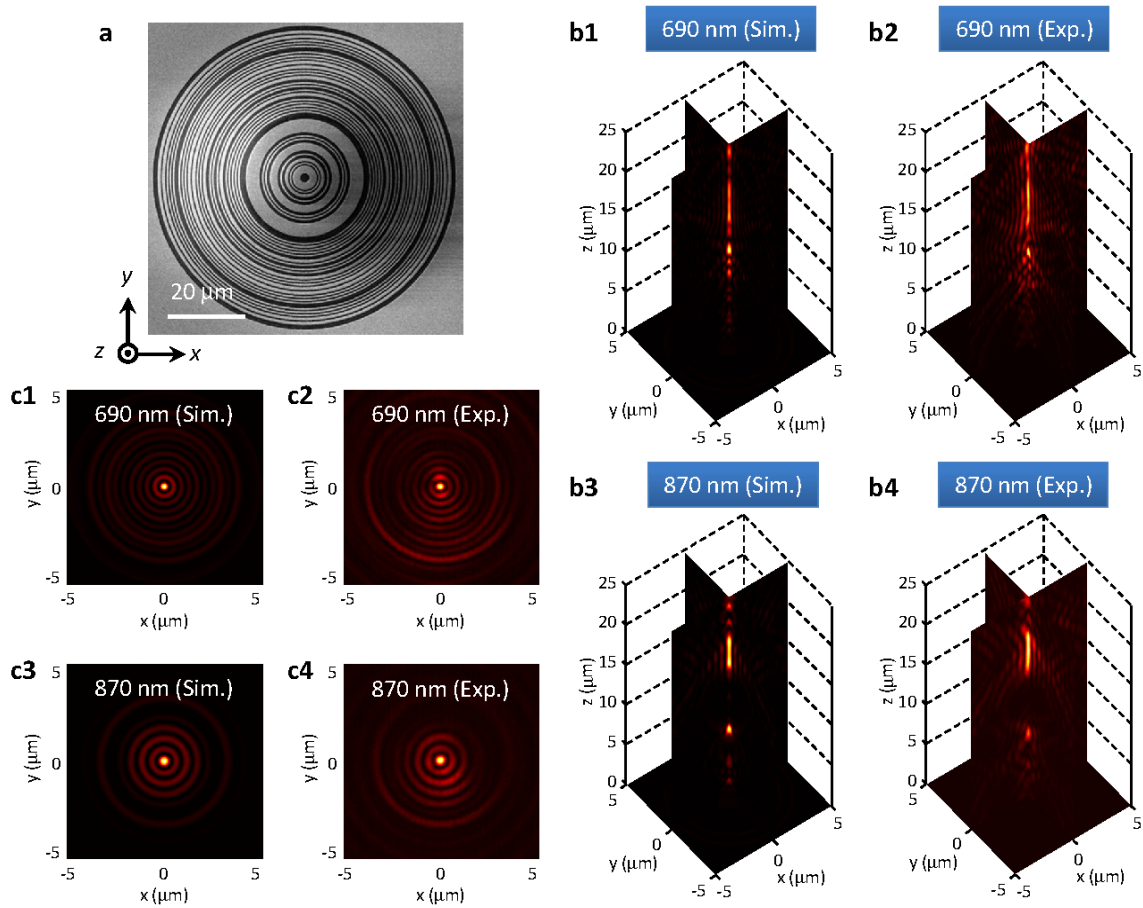

**Figure 4** Visible achromatic amplitude mask SOL. (**a**) SEM micrograph of the fabricated mask with diameter of 80 µm and smallest ring width of 400 nm on a 100 nm-thick gold film. (**b**) Experimental and simulated diffraction patterns in the *xz* and *yz* cross-sections, where the achromatic super-oscillatory hotspots are generated at $z_f$=18 µm for $\lambda_1$=690 nm and $\lambda_2$=870 nm. (**c**) Corresponding field patterns in the achromatic transverse focal plane. Incident beam is *x*-polarized.

The mask is shown in Fig. 4a. Simulated field patterns in the central cross-sections along (*xz* plane) and perpendicular to incident polarization (*yz* plane) are shown in Fig. 4b. The incident beam is an *x*-polarized plane wave. Achromatic performance is centered at $z_f$ =18 µm, where elongated hotspots at $\lambda_1$=690 nm (15 µm to 20.1 µm) overlaps with hotspot at $\lambda_2$=870 nm (16 µm to 19.9 µm) for $\lambda_2$. The FWHM of the hotspot is 280 nm (0.405×$\lambda_1$) and 367 nm (0.422×$\lambda_2$) for $\lambda_1$ and $\lambda_2$ respectively and is sub-diffraction-limited. For a conventional lens with of this size the diffraction-limited hotspot will be 378 nm at $\lambda_1$=690 nm and 477 nm at $\lambda_2$=870 nm.

The experimentally measured field structures are presented in Fig. 4b2 and Fig. 4b4. Hotspots with comparable depth of focus are generated at $z_f$ =18 µm, in agreement with the theoretical predictions. At $\lambda_1$=690 nm the observed dimensions of the hot-spots were 0.42×$\lambda_1$ for the direction along the incident polarization and 0.46×$\lambda_1$ for perpendicular direction. For $\lambda_2$=870 nm the spot size was 0.43×$\lambda_2$ and 0.44×$\lambda_2$ in two directions respectively, see Fig. 4c2 and Fig. 4c4. The fraction of energy focused into the sub-wavelength hotspots at the target plane at 18 µm was found to be 0.47% and 1% for $\lambda_1$ and $\lambda_2$, respectively.

Here we note that the conventional binary Fresnel zone plate is highly dispersive and its focal distance will depend on the wavelength: progressively shifting away from the lens for longer wavelengths. Thus Fresnel zone plate cannot offer an achromatic solution (See Supplementary Information-Section 4).

**RGB 'white'-light apochromatic amplitude mask SOL**

We found that designing a single-element lens that focuses light across the entire visible range into a single hotspot is difficult (see the previous SOL performance for the intermediate wavelengths between $\lambda_1$ and $\lambda_2$ in Supplementary Information-Section 5). However, we have been able to design a SOL which simultaneously focuses the three primary colors of the additive model that together can produce perceived white light. We have developed an apochromatic SOL for red ($\lambda_R$=633 nm), green ($\lambda_G$=532 nm) and blue ($\lambda_B$=405 nm) light, as illustrated in Fig. 5a. It is well know from computer display and digital imaging technology, that any perceived color can be generated by combining these three colors.

Figure 5b shows an SEM micrograph of the mask. We target a working distance of 10 µm and a FWHM of 0.4×$\lambda$. The simulated intensity profiles in the longitudinal cross-section are given in Fig. 5c. The corresponding experimental data are shown in Fig. 5d. The field structures in the focal plane for R, G, B wavelengths are displayed in Figs. 5e1-5e3. The simulated hotspot sizes were 0.438×$\lambda_R$ at $\lambda_R$=633 nm, 0.422×$\lambda_G$ at $\lambda_G$=532 nm and 0.523×$\lambda_B$ at $\lambda_B$=405 nm. Experimentally we obtained hotspots of (0.457±0.01)×$\lambda_R$, (0.445±0.04)×$\lambda_G$ and (0.54±0.03)×$\lambda_B$ respectively. The fraction of energy focused into the sub-wavelength hotspots at the target plane was found to be 1.35%, 1.18% and 1.54% for $\lambda_R$, $\lambda_G$, and $\lambda_B$ respectively.

As expected we were able to acquire a 'white' super-oscillatory hotspot by focusing supercontinuum fiber laser radiation of which was spectrally conditioned at the selected RGB wavelengths by a programmable acousto-optic tunable filter, see Fig. 5e4.

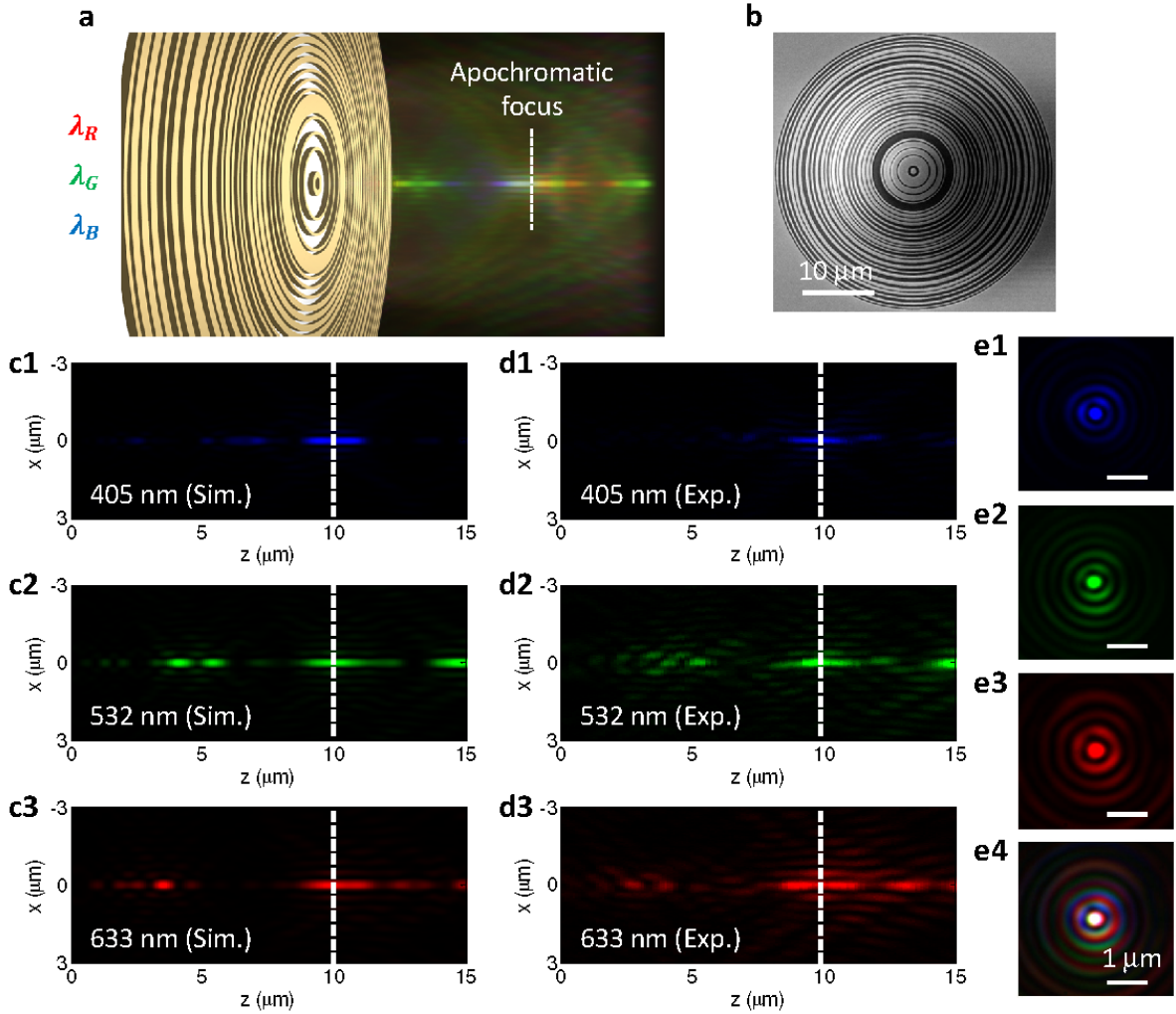

**Figure 5** RGB 'white'-light apochromatic amplitude mask SOL. (**a**) Apochromatic SOL focuses simultaneously at three different wavelengths, red ($\lambda_R$=633 nm), green ($\lambda_G$=532 nm) and blue ($\lambda_B$=405 nm) that can form a 'white' super-oscillatory hotspot. (**b**) SEM micrograph of the fabricated mask with diameter of 40 µm, and working distance of 10 µm. (**c**)(**d**) Simulated (c) and experimental (d) diffraction patterns in the *xz* cross-section. The vertical dashed white lines indicate the focal plane. (**e**) Experimentally registered intensity patterns in the transverse focal plane: (e1) for $\lambda_B$, (e2) for $\lambda_G$, (e3) for $\lambda_R$, and (e4) for RGB wavelengths by simultaneously switching on the three channels. Images are taken by a color CCD camera.

## CONCLUSIONS

We experimentally demonstrated that chromatic aberration and the optical diffraction limit can be simultaneously overcome by SOL consisting of a single planar optical element. The single-element SOL creates sub-diffraction hotspots by delicate constructive interference of propagating optical waves with different wavevectors, and therefore allows foci at different wavelengths to spatially overlap in achromatic fashion.

Sub-wavelength focusing of SOLs comes at a price. Only small fraction of light incident on the lens is actually focused in the central hotspot while the remains are mainly distributed in broad "halo" rings around the hotspot. For some applications the effect of halo can be eliminated. For instance, in imaging light scattered by halo can be suppressed using a confocal technique[30] while in the heat assisted magnetic recording application it can be illuminated by using additional thin apertures[46]. In some nonlinear optical applications such as coherent anti-Stokes Raman spectroscopy (CARS)[47] suppression of the halo comes naturally through the nonlinear interactions at different frequencies.

In the focusing devices described here 1% to 3% of incident energy is focused in the central hotspot, while the intensity level could be a factor of 3 higher than the intensity level of the incident wave onto the lens, as for instance, in the fiberized lens described above. The low level of through-put efficiency may be prohibitive for some demanding application, but can be tolerable for others. For instance, a low-loss dielectric SOL that is 40 µm in diameter will be able to sustain incident CW laser radiation of a few watts, corresponding to focal intensities of a few hundred $kW/cm^2$. When used with pulsed lasers, the lens will be able to sustain a few hundred of $mJ/cm^2$ of incident fluence with femtosecond and picosecond pulses and up to a few $J/cm^2$ with nanosecond pulses[48,49]. It will deliver a similar level of energy density in the hotspot. Such levels of optical excitations will be overwhelmingly sufficient for raster imaging, lithography and nanofabrication with light.

We realized metallic and dielectric binary amplitude and phase masks that are wavelength scalable and can work for different spectral bands. In the infrared band we implemented achromatic SOLs at the tip of a large-mode-area single-mode photonic crystal fibers and silicon wafer. For the visible band, we developed an achromatic SOL for two well-separated wavelengths and an apochromatic SOL for RGB wavelengths suitable for creating a sub-diffraction 'white' hotspot.

We anticipate that the achromatic sub-diffraction photonic focusing devices could serve as super-resolved focusing and imaging tools for a broad range applications in nonlinear optics, photography, retinal diagnostics[50], low-cost fiberized microscopy, non-invasive and label-free biological imaging, pump-probe experiment for ultrafast dynamics, excitation and collection of photoluminescence, coherent anti-Stokes Raman scattering for super-resolution bio-imaging, nonlinear imaging and nanofabrication using two/three-photon generation.

1   Wang Y, Yun W, Jacobsen C. Achromatic Fresnel optics for wideband extreme-ultraviolet and X-ray imaging. *Nature* 2003; **424**: 50–53.
2   Greisukh GI, Ezhov EG, Stepanov SA. Diffractive–refractive hybrid corrector for achro- and apochromatic corrections of optical systems. *Appl Opt* 2006; **45**: 6137–6141.
3   Mikš A, Novák J. Method for primary design of superachromats. *Appl Opt* 2013; **52**: 6868–6876.
4   Lin D, Fan P, Hasman E, Brongersma ML. Dielectric gradient metasurface optical elements. *Science* 2014; **345**: 298–302.
5   Verslegers L, Catrysse PB, Yu Z, White JS, Barnard ES *et al*. Planar lenses based on nanoscale slit arrays in a metallic film. *Nano Lett* 2008; **9**: 235–238.
6   Ishii S, Shalaev VM, Kildishev AV. Holey-metal lenses: sieving single modes with proper phases. *Nano Lett* 2012; **13**: 159–163.

7   Chen XZ, Huang LL, Mühlenbernd H, Li GX, Bai BF *et al*. Dual-polarity plasmonic metalens for visible light. *Nat Commun* 2012; **3**: 1198.
8   Aieta F, Genevet P, Kats MA, Yu NF, Blanchard R *et al*. Aberration-free ultrathin flat lenses and axicons at telecom wavelengths based on plasmonic metasurfaces. *Nano Lett* 2012; **12**: 4932–4936.
9   Ni X, Ishii S, Kildishev AV, Shalaev VM. Ultra-thin, planar, Babinet-inverted plasmonic metalenses. *Light Sci Appl* 2013; **2**: e72.
10  Pors A, Nielsen MG, Eriksen RL, Bozhevolnyi SI. Broadband focusing flat mirrors based on plasmonic gradient metasurfaces. *Nano Lett* 2013; **13**: 829–834.
11  Fattal D, Li J, Peng Z, Fiorentino M, Beausoleil RG. Flat dielectric grating reflectors with focusing abilities. *Nature Photon* 2010; **4**: 466–470.
12  Arbabi A, Horie Y, Ball AJ, Bagheri M, Faraon A. Subwavelength-thick lenses with high numerical aperture and large efficiency based on high-contrast transmitarrays. *Nat Commun* 2015; **6**: 7069.
13  Aieta F, Kats MA, Genevet P, Capasso F. Multiwavelength achromatic metasurfaces by dispersive phase compensation. *Science* 2015; **347**: 1342–1345.
14  Arbabi E, Arbabi A, Kamali SM, Horie Y, Faraon A. Multiwavelength polarization-insensitive lenses based on dielectric metasurfaces with meta-molecules. *Optica* 2016; **3**: 628.
15  Zhao ZY, Pu M, Gao H, Jin J, Li X *et al*. Multispectral optical metasurfaces enabled by achromatic phase transition. *Sci Rep* 2015; **5**: 15781.
16  Wang P, Mohammad N, Menon R. Chromatic-aberration-corrected diffractive lenses for ultra-broadband focusing. *Sci Rep* 2016; **6**: 21545.
17  Hein B, Willig KI, Hell SW. Stimulated emission depletion (STED) nanoscopy of a fluorescent protein-labeled organelle inside a living cell. *Proc Natl Acad Sci USA* 2008; **105**: 14271–14276.
18  Betzig E, Patterson GH, Sougrat R, Lindwasser OW, Olenych S *et al*. Imaging intracellular fluorescent proteins at nanometer resolution. *Science* 2006; **313**: 1642–1645.
19  Rust MJ, Bates M, Zhuang X. Sub-diffraction-limit imaging by stochastic optical reconstruction microscopy (STORM). *Nat Methods* 2006; **3**: 793–795.
20  Pendry JB. Negative refraction makes a perfect lens. *Phys Rev Lett* 2000; **85**: 3966–3969.
21  Zhang X, Liu ZW. Superlenses to overcome the diffraction limit. *Nature Mater* 2008; **7**: 435-441.
22  Liu ZW, Durant S, Lee H, Pikus Y, Fang N *et al*. Far-field optical superlens. *Nano Lett* 2007; **7**: 403–408.
23  Rho J, Ye Z, Xiong Y, Yin X, Liu Z *et al*. Spherical hyperlens for two-dimensional sub-diffractional imaging at visible frequencies. *Nat Commun* 2010; **1**: 143.
24  Toraldo di Francia G. Supergain antennas and optical resolving power. *Nuovo Cimento Suppl* 1952; **9**: 426–435.
25  Sales TRM, Morris GM. Diffractive superresolution elements. *J Opt Soc Am A* 1997; **14**: 1637–1646.
26  Sheppard CJR, Calvert G, Wheatland M. Focal distribution for superresolving toraldo filters. *J Opt Soc Am A* 1998; **15**: 849–856.
27  Zhu H, Xu D, Wang X, Liu W, Yun M *et al*. Quasi-achromatic superresolving phase only pupil filters. *Opt Commun* 2012; **285**: 5062–5067.
28  Zheludev NI. What diffraction limit? *Nature Mater* 2008; **7**: 420–422.
29  Huang FM, Zheludev NI. Super-resolution without evanescent waves. *Nano Lett* 2009; **9**: 1249–1254.
30  Rogers ETF, Lindberg J, Roy T, Savo S, Chad JE *et al*. A super-oscillatory lens optical microscope for subwavelength imaging. *Nature Mater* 2012; **11**: 432–435.
31  Rogers ETF, Zheludev NI. Optical super-oscillations: sub-wavelength light focusing and super-resolution imaging. *J Opt* 2013; **15**: 094008.
32  Roy T, Rogers ETF, Zheludev NI. Sub-wavelength focusing meta-lens. *Opt Express* 2013; **21**: 7577–7582.
33  Yuan GH, Rogers ETF, Roy T, Adamo G, Shen ZX *et al*. Planar super-oscillatory lens for sub-diffraction circular polarization optical needles at violet wavelengths. *Sci Rep* 2014; **4**: 6333.

34  Huang K, Liu H, Garcia-Vidal FJ, Hong M, Lukyanchuk B *et al*. Ultrahigh-capacity non-periodic photon sieves operating in visible light. *Nat Commun* 2015; **6**: 7059.
35  Wang Q, Rogers ETF, Gholipour B, Wang CM, Yuan GH *et al*. Optically reconfigurable metasurfaces and photonic devices based on phase change materials. *Nature Photon* 2016; **10**: 60–65.
36  Yuan GH, Vezzoli S, Altuzarra C, Rogers ETF, Couteau C *et al*. Quantum super-oscillation of a single photon. *Light Sci Appl* 2016; **6**: e16127.
37  Wong AMH, Eleftheriades GV. Advances in imaging beyond the diffraction limit. *IEEE Photon J* 2012; **4**: 586–589.
38  Wong AMH, Eleftheriades GV. An optical super-microscope for far-field, real-time imaging beyond the diffraction limit. *Sci Rep* 2013; **3**: 1715.
39  Aharonov Y, Colombo F, Sabadini I, Struppa DC, Tollaksen J. Some mathematical properties of superoscillations. *J Phys A Math Theor* 2011; **44**: 365304.
40  Ferreira PJSG, Kempf A, Reis MJCS. Construction of Aharonov-Berry's super-oscillations. *J Phys A Math Theor* 2007; **40**: 5141–5147.
41  Berry MV, Popescu S. Evolution of quantum superoscillations and optical superresolution without evanescent waves. *J Phys A Math Gen* 2006; **39**: 6965–6977.
42  Berry MV, Moiseyev N. Superoscillations and supershifts in phase space: Wigner and Husimi function interpretations. *J Phys A Math Theor* 2014; **47**: 315203.
43  Rogers ETF, Savo S, Lindberg J, Roy T, Dennis M, *et al*. Super-oscillatory optical needle. *Appl Phys Lett* 2013; **102**: 031108.
44  Qin F, Huang K, Wu J, Jiao J, Luo X *et al*. Shaping a subwavelength needle with ultra-long focal length by focusing azimuthally polarized light. *Sci Rep* 2015; **5**: 09977.
45  Jin N, Rahmat-Samii Y. Advances in particle swarm optimization for antenna designs: Real-number, binary, single-objective and multiobjective implementations. *IEEE Trans Antenn Propag* 2007; **55**: 556–567.
46  Yuan GH, Rogers ETF, Roy T, Shen Z, Zheludev NI. Flat super-oscillatory lens for heat-assisted magnetic recording with sub-50 nm resolution. *Opt Express* 2014; **22**: 6428–6437.
47  Begley RF, Harvey AB, Byer RL. Coherent anti-Stokes Raman spectroscopy. *Appl Phys Lett* 1974; **25**: 387–390.
48  Bonse J, Baudach S, Krüger J, Kautek W, Lenzner M. Femtosecond laser ablation of silicon-modification thresholds and morphology. *Appl Phys A* 2002; **74**: 19–25.
49  Wang X, Shen ZH, Lu J, Ni XW. Laser-induced damage thresholds of silicon in millisecond, nanosecond, and picosecond regimes. *J Appl Phys* 2010; **108**: 033103.
50  Godara P, Dubis AM, Roorda A, Duncan JL, Carroll J. Adaptive optics retinal imaging: emerging clinical applications. *Optom Vis Sci* 2010; **87**: 930–941.

## ACKNOWLEDGEMENTS

This work was supported by the Agency for Science, Technology and Research (A*STAR) of Singapore (Grants 122-360-0009), the Singapore Ministry of Education (Grant MOE2011-T3-1-005), the Engineering and Physical Sciences Research Council UK (Grants EP/F040644/1 and EP/M009122/1) and the University of Southampton Enterprise Fund. The authors thank X. H. Li and L. B. Wang for assistance with fibre cleaving, G. Adamo and H. Krishnamoorthy for ellipsometry measurement of silicon wafer, and P. Smith for fruitful discussions. The data from this paper can be obtained from the University of Southampton ePrints research repository: http://dx.doi.org/10.5258/SOTON/394664.

# Supplementary Information for

# Achromatic super-oscillatory lenses with sub-wavelength focusing

Guang Hui Yuan[1#], Edward T. F. Rogers[2,3], and Nikolay I. Zheludev[1,2*]

[1]*Centre for Disruptive Photonic Technologies, The Photonic Institute, SPMS, Nanyang Technological University, Singapore 637371, Singapore*

[2]*Optoelectronics Research Centre and Centre for Photonic Metamaterials, University of Southampton, Highfield, Southampton, SO17 1BJ, UK*

[3]*Instiute for Life Sciences, University of Southampton, Highfield, Southampton, SO17 1BJ, UK*

[*]nzheludev@ntu.edu.sg and [#]ghyuan@ntu.edu.sg

**Contents:**

1. Design parameters of achromatic SOLs
2. Experimental setup for SOL characterization
3. Optical refractive index of silicon wafer
4. Comparison between achromatic SOL and binary Fresnel zone plates
5. Evolution of diffraction patterns at intermediate wavelengths

- **Design parameters of achromatic SOLs**

All design parameters of the achromatic SOLs discussed in this paper are shown in Table S1, including working wavelengths, materials and substrates, working distance, size and dimensions

| Table S1. Design parameters of the achromatic SOLs ||||||||
|---|---|---|---|---|---|---|---|
| Sample # | Wavelengths (nm) | Material and substrates | Working distance ($\mu m$) | SOL size | Dimension ($N$) | Amp/Phase modulation | Binary values (from centre to outward) * |
| 1. Fiberized SOL | 1300/1550 | 100 nm Au on fiber tip | 8 | Dia. 30 $\mu m$ | 75 | Amplitude | 000000000000000000000000001000001100100110111010011111000011110000001101111 |
| 2. Dielectric SOL | 1300/1550 | Silicon wafer | 20 | Dia. 80 $\mu m$ | 50 | Phase | 11000100010110101000000110010000110010101011010 |
| 3. Visible SOL | 690/870 | 100 nm Au on silica glass | 18 | Dia. 80 $\mu m$ | 100 | Amplitude | 1110000010100110110000011010110000000011110101010110010110101100101010010010101011101010100100111 |
| 4. RGB SOL | 405/532/633 | 100 nm Au on silica glass | 10 | Dia. 40 $\mu m$ | 100 | Amplitude | 0011000000010000100001011111101010010010011011010110010100101010011001100110110011001100110110110100 |
| Remark | *The binary values 0 and 1 correspond to zero and unit transmittance for amplitude masks, and two-level phase for phase masks. In the dielectric SOL, the binary phase is $0/1.2\pi$ for 1300 nm and $0/\pi$ for 1550 nm. Both designs take the material dispersion into account during the mask optimization procedure. |||||||

Figure S1 gives an example of the fiber sample before and after SOL integration, where the hexagonally arranged air holes with a central defect confine the optical mode and a well-aligned and concentrically symmetric multi-ring structure can be clearly seen in the core region.

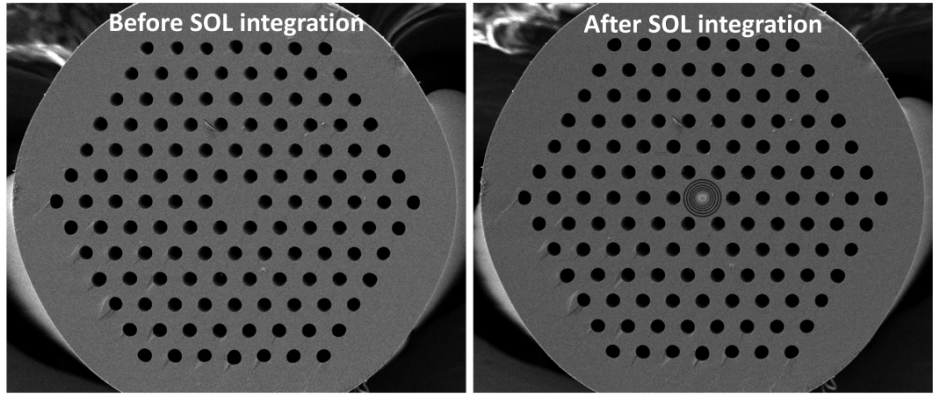

**Figure S1.** Large-mode-area single-mode photonic crystal fiber before (left) and after (right) SOL integration. The fiber end surface is coated with a 100 nm-thick gold film which serves as an amplitude mask.

- **Experimental setup for SOL characterization**

The experimental setup is sketched in Fig. S2. All the experiments were conducted using a super-continuum laser source (Fianium WL-SC-400-8, UK) with spectral range from 400 nm to 2500 nm. The individual wavelength channel can be selected and turned on/off using compatible acousto-optic tunable filters (AOTF) and with the relative power in each channel controllable by software. After collimation by a fiber coupler (FC1), the beam is directed into a customized dual-mode microscope (Nikon Eclipse Ti-E/LV) using a pair of mirrors (M1, M2) and illuminates the SOLs after reflection by a dichroic mirror (DM) orientated at 45°. Since the super-oscillatory fields are formed by delicate interference of propagating waves, they can be mapped into the far-field and directly imaged by a conventional optical imaging system. We used a high-magnification, high-numerical aperture objective (Nikon CFI LU Plan APO EPI 150X, NA=0.95) to collect the diffracted fields which are additionally magnified by a 4X magnification changer (Nikon C-Mount TV Adapter VM 4X) and subsequently imaged by a suitable high-resolution camera (Andor Neo sCMOS camera for 690 nm/870 nm; Thorlabs USB2.0 color CMOS camera for RGB wavelengths; Photonic Science SWIR InGaAs VGA camera for IR wavelengths). The transverse cross-section distributions at specific propagation distances are obtained by z-scanning of the SOL mounted on a 3-axis piezo stage (PI P545). The longitudinal cross-section distributions are captured simultaneously from real-time data processing in Labview.

When measuring the fiberized SOL, we simply mount the fiber onto the same sample stage and use similar characterization procedures. The second fiber coupler (FC2) is used to couple the laser beam into a photonic crystal fiber (PCF, ~one-meter) on the other end of which is the SOL.

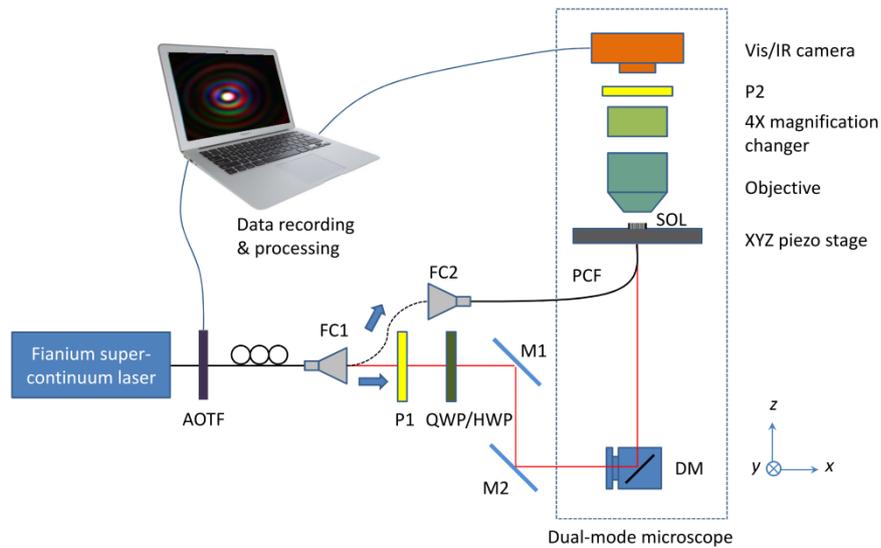

**Figure S2.** Experimental setup for SOL characterization. AOTF: acousto-optic tunable filter; P1, P2: polarizer; M1, M2: optical mirrors; FC1, FC2: fiber couplers; PCF: photonic crystal fiber; DM, dichroic mirror.

- **Optical refractive index of silicon wafer**

The optical properties of the silicon wafer (University Wafer Inc., DSP, 100, p-type) are measured by a microspectrophotometer (Jasco MSV-5000 series). By taking the transmittance and reflectance data and using data-fitting, the complex refractive index can be retrieved as shown in Fig. S3, where the real part of the refractive index at wavelengths of 1300 nm and 1550 nm is $n_{1300}$=3.509 and $n_{1550}$=3.481, while the imaginary part (related to absorption) is extremely small and can be neglected.

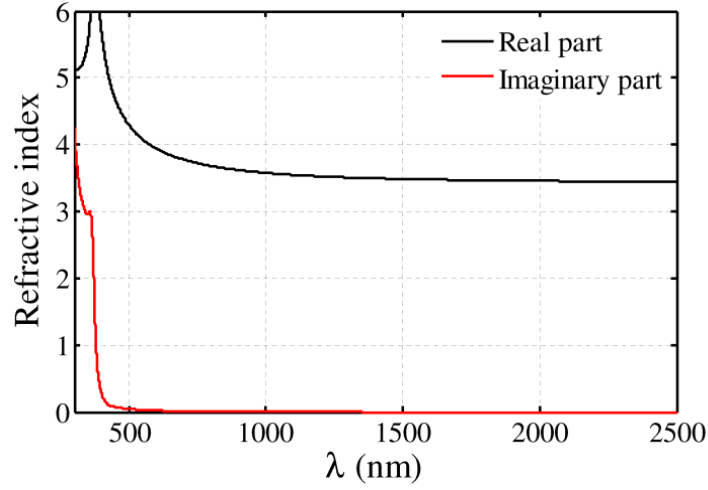

**Figure S3.** Experimentally measured refractive index of silicon wafer.

- **Comparison between achromatic SOL and binary Fresnel zone plates**

Compared with our achromatic SOL, a conventional diffractive optical element like a binary Fresnel zone plate is highly dispersive and will focus different wavelengths into different axial positions. We give examples in Fig. S4. The radii of alternating transparent and opaque zones are given by the standard formula:

$$r_n = \sqrt{n\lambda f + \frac{n^2\lambda^2}{4}}, n = 1,2,3,\ldots\ldots \quad (1)$$

where $\lambda$ and $f$ are the working wavelength and focal length respectively.

For an FZP working at $\lambda_1$=690 nm and $f$=18 $\mu$m, we used 75 zones in the design and the SEM image is shown in Fig. S4(a) where the fabrication is conducted on a 100nm-thick gold film on a glass substrate. Its focusing performance at $\lambda_1$=690 nm and $\lambda_2$=870 nm is characterized by both angular spectrum simulation and experiment which agree well with each other, as shown in Fig. S4(b). Indeed, the focal spot for $\lambda_1$ is located at $z$=18 $\mu$m as predicted, but for $\lambda_2$ hotspot shifts much closer to the FZP and no hotspot can be observed at $z$=18 $\mu$m. Similarly, for the fabricated 59-zone FZP working for $\lambda_2$=870 nm and $f$=18 $\mu$m as shown in Fig. S4(c), the focal spot is formed at $z$=18 $\mu$m for $\lambda_2$, but shifts to be farther away from the FZP for shorter wavelength $\lambda_1$, as seen in Fig. S4(d). The theoretical energy concentration ratio at $z$=18 $\mu$m is 13.8% and 13.6% respectively, for the two design wavelengths, which is much higher than that of achromatic SOL with similar structural parameters in terms of mask size and focal length, as discussed in the main text. When not used at the design wavelength, the energy concentration ratio for the two FZPs is 0.4% and 0.14%. While the efficiency of the SOL is worse than an FZP working at its design wavelength (an inevitable consequence of super-oscillatory focusing), it has noticeably better efficiency at the second wavelength of interest.

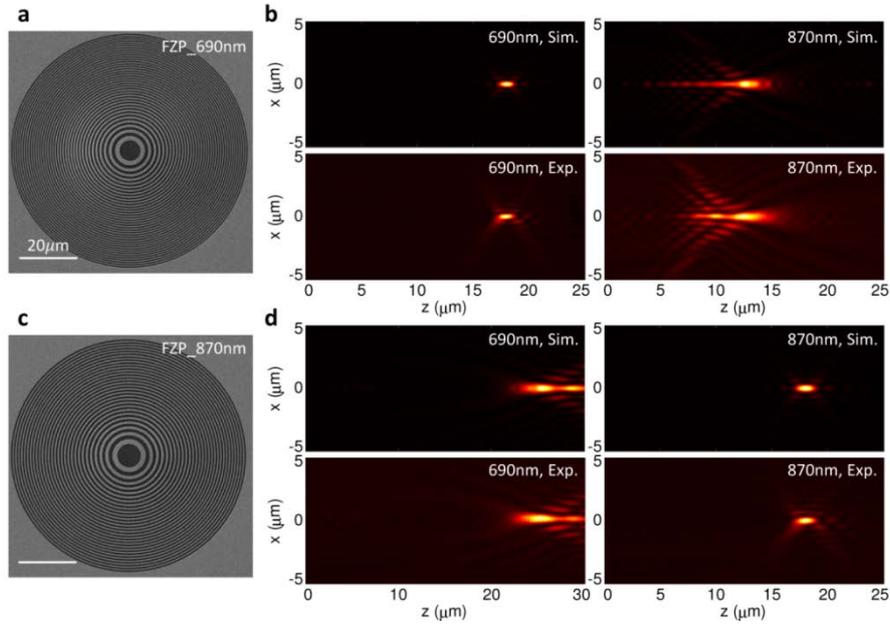

**Figure S4. a,** SEM image of the FZP designed for $\lambda_1$=690 nm and focal length of 18 $\mu$m. **b,** Simulation calculated (top) and experimentally measured (bottom) diffraction patterns in the longitudinal cross-sections for $\lambda_1$=690 nm and $\lambda_2$=870 nm. **c,** SEM image of the FZP with same focal length of 18 $\mu$m for $\lambda_2$=870 nm. **d,** Simulation calculated (top) and experimentally measured (bottom) diffraction patterns in the longitudinal cross-sections for $\lambda_1$=690 nm and $\lambda_2$=870 nm.

- **Evolution of diffraction patterns at intermediate wavelengths**

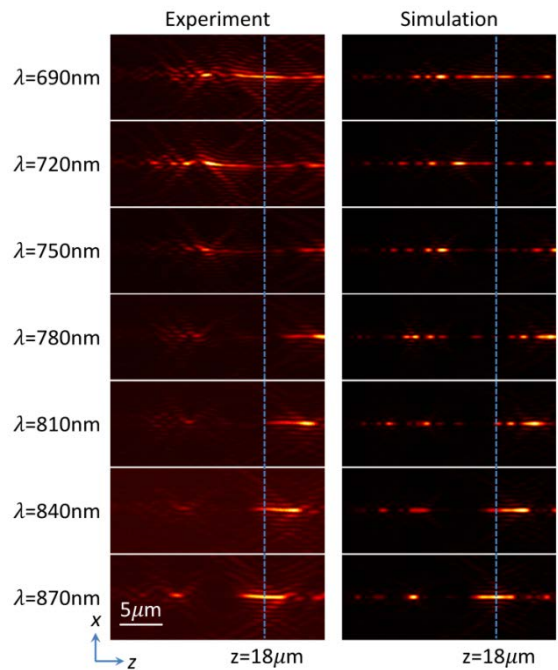

**Figure S5.** Longitudinal cross-section diffraction patterns at tunable wavelengths from 690 nm to 870 nm with a step of 30 nm. (Left column) Experimental data. (Right column) Simulation data obtained from angular spectrum calculations. Good agreement is found between overall profiles and evolution dynamics of the hotspots. The vertical dashed blue lines indicate the designed achromatic focal plane.